\def\b{\begin{equation}}
\def\e{\end{equation}}
\def\l{\left}
\def\r{\right}
\documentclass[prl,aps,twocolumn, showpacs, showkeys, floats]{revtex4}
\begin{document}

\title{Why Gravitational Contraction Must be Accompanied by Emission of Radiation both
in Newtonian and Einstein
Gravity}

\author{Abhas Mitra}

\email {Abhas.Mitra@mpi-hd.mpg.de}
\affiliation { MPI fur Kernphysik, Saupfercheckweg 1, D-69117
Heidelberg, Germany}

 \altaffiliation[Also at ]{Theoretical Astrophysics Section, Bhabha Atomic
Research Center, Mumbai-400085, India }



\date{\today}
\begin{abstract}
By using virial theorem, Helmholtz and Kelvin showed that the
contraction of a bound self-gravitating system must be accompanied
by release of radiation energy {\em irrespective of the details}  of
the contraction process. This happens because the total Newtonian
energy of the system $E_N$ (and not just the Newtonian gravitational
potential energy $E_g^N$) decreases for such contraction. In the era
of General Relativity (GR) too, it is justifiably believed that
gravitational contraction must release radiation energy. However no
GR version of (Newtonian) Helmholtz- Kelvin (HK) process  has ever
been derived.   Here, for the first time, we derive the GR version
of  the appropriate virial theorem and Helmholtz Kelvin mechanism by
simply equating the well known expressions for the {\sl
gravitational mass}  and the {\sl Inertial Mass}  of a spherically
symmetric static fluid. Simultaneously, we  show that  the GR
counterparts of global ``internal energy'', ``gravitational
potential energy'' and ``binding energy'' are actually different
from what have been used so far.  Existence of this GR HK process
asserts that, in Einstein gravity too, gravitational collapse must
be accompanied by emission of radiation irrespective of the {\em
details of the collapse process}.

\end{abstract}


\maketitle


\section{Introduction}
It is generally believed that General Relativistic (GR)
gravitational collapse must be accompanied by the emission of
radiation such as photons and neutrinos. There have been many
studies on radiative spherical GR collapse and we may recall here
only a few of them\cite{1}. Though all such studies vastly differ in
their details, nevertheless, all of them indicate that the fluid
becomes hotter during the collapse while net amount of radiated
energy steadily increases. Surprisingly, though, such effects seem
obvious at  first sight, one would wonder whether there is a
fundamental reason for occurrence of such physical effects during
gravitational collapse. Put another way, whether, irrespective of
solution of actual collapse equations with their associated
assumptions and simplifications,  one can demand from a general
perspective, that actual gravitational collapse must be accompanied
by both emission of radiation and heating up of the fluid. From
thermodynamics perspective, one can ask whether the phenomenon of
occurrence ``negative specific heat'', known for Newtonian gravity
for a long time,  must be valid in Einstein gravity too. To
appreciate this, let us recall that the specific heat is defined
through $C = dQ/dT$ where $dQ$ is the amount of heat injected into
the system and $dT$ is the corresponding increment of temperature.
Gravitational compression raises the temperature so that $dT >0$. A
negative $C$ would then demand $dQ <0$ and vice-versa. Hence a
negative $dQ$ means  loss of heat (radiation) from the system and
vice-versa.

We emphasize here that, in a strict sense,  this phenomenon of
``negative specific heat'' is known only for weak Newtonian gravity.
Intuitively such an effect is expected to be more pronounced for a
fluid subject to much stronger Einstein gravity. But the actual fact
is that, there is no proper GR theorem which can assert that global
Einstein gravity too is marked by the same ``negative specific
heat''.

To highlight this, in Sec 1, we will first review  the case in the
Newtonian gravitation. This would show why Newtonian collapse must
be accompanied by radiation howsoever small it may be and why global
gravitation is characterized by ``negative specific heat'' in
Newtonian case. Then it would be emphasized that a corresponding GR
derivation is non existent and accordingly we shall present an exact
GR counterpart of this Newtonian process. We would then
automatically arrive at global definitions GR Self-Gravitational
energy and Binding Energy from the perspective of global energy
conservation of a static spherically symmetric fluid. The entire
exercise will show why, irrespective of the details, GR collapse
must be accompanied by emission of radiation and an increasing fluid
temperature.

\section{Newtonian Gravitational Collapse}

 As per
Chandrasekhar\cite{2},  von Helmholtz first proposed in 1854 that
contraction of self-gravitating bodies should emit radiation. Few
years later, in 1861,  Kelvin\cite{3} elaborated on this process of
energy generation which may be called as Helmholtz - Kelvin (HK)
process. Without going into further historical account, the basic
physics behind the H-K process is reviewed below from a relatively
modern perspective\cite{4,5}:

If we consider a spherically symmetrical static isotropic fluid in hydrodynamical equilibrium, it follows
that

\b
E_g^N + 3\int p~ dV = 4 \pi R^3 ~p_b
\e
where, $E_g^N$ is the Newtonian gravitational potential energy, $p = p(r)$ is the isotropic
pressure, $p_b$ is the pressure at the boundary $r=R$, and the volume element $dV = 4 \pi r^2
dr$. For a laboratory gas sphere, it is possible to have $p(r)\approx uniform \approx p_b$. In such a case, Eq.(1) would reduce to
\b
E_g^N + 4 \pi R^3 ~p_b \approx 4\pi R^3~p_b
\e
and which would express the obvious fact that for a laboratory gas $E_g^N \approx 0$.
Note that it is necessary to have $p_b >0$ in such a case.

In contrast, if the fluid is assumed to be self-contained, i.e., {\em bound} by its own gravity
then one expects
\b
p_b =0
\e
and  obtains the better known form of Eq.(1):

\b E_g^N + 3\int p~ dV = 0 \e This is known as static and scalar
virial theorem. If the adiabatic index of the fluid is $\gamma$,
then \b p = (\gamma -1) e \e where $e$ is the  internal energy
density. For some ideal fluids, $\gamma$ may also be considered as
the ``ratio of specific heats''. For example, for a monoatomic ideal
gas having an equation of state (EOS)  $p = n k T$, where $n$ is
number density of the monoatomic molecules and $k$ is the Boltzmann
constant $\gamma$ is the ratio of specific heats with a unique value
$\gamma \equiv 5/3$. For an ultrarelativistic gas with particle
momenta $\to \infty$ or for a pure photon gas, $\gamma$ is again the
ratio of specific heats having the unique value $\gamma \equiv
 4/3$\cite{2}. The internal energy for the entire fluid is \b U =
\int e ~dV \e Using Eqs.(5) and (6) in (4), and assuming $\gamma$ to
be uniform, we have \b E_g^N + 3(\gamma -1)~ U = 0 \e so that \b U =
{-1 \over3(\gamma -1)} E_g^N \e The total Newtonian energy of the
fluid is \b E_N = E_g^N + U = {3 \gamma -4 \over3(\gamma -1)} ~
E_g^N \e Note that for the above mentioned ideal gas having an EOS
$p = n k T$,  Eq.(7) reduces to the more familiar form $E_g^N + 2 U
=0$ because here $\gamma = 5/3$. However Eqs.(1-9) are valid for any
fluid obeying relation (5) and not necessarily by an ``ideal gas''
alone, as long as the fluid may be assumed to obey an EOS of the
form (5). Thus, in principle, $\gamma$ is arbitrary here subject to
general thermodynamical constraints.

 In  Eq.(9), $E_N$ is the Newtonian binding energy of the
fluid and must be negative for a fluid which is already assumed to
be {\em bound}. For attractive gravity, for {\em any case, bound or
unbound}, one must have $E_g^N <0$. In the present case, as soon as
we set $p_b=0$, we imply the system to be self-bound, and one must
have $E_N <0$. Thus from Eq.(9), it transpires that, one must have
$\gamma > 4/3$. A limiting case of $\gamma =4/3$ would signify a
transition to unbound systems. For ``unbound systems'' one expects
to have, $E_N >0$. If the system would be unbound, one would  have
$p_b >0$ and further the system would not be in hydrostatic
equilibrium. In such a case, one needs to use dynamical form of
virial theorem to study it.    Also, by noting Eq.(7), one might
think that for unbound systems, one might have $\gamma <4/3$. But
this would be an incorrect conclusion, because as explained above,
for unbound systems, Eq.(7) would cease to be valid. It may be borne
in mind that $\gamma$ is an inherent thermodynamical parameter and
cannot be dictated by gross global hydrodynamical behavior of the
fluid.

A limiting value of $\gamma=4/3$ corresponds to a situation when the
momenta of the constituent particles of the fluid, $P =\infty$, and
thus, cannot be strictly realized except for singular
situations\cite{6}. For instance note that the critical
ultrarelativistic White Dwarf of Chandrasekhar has $R=0$ because it
strictly corresponds to a fluid having $\gamma =4/3$\cite{2}.

If,  {\em additionally}, the fluid obeys a polytropic equation of
state \b p = K \rho^{\gamma_p} \e where, $K$ and $\gamma_p$ are
uniform over the fluid, it follows that \b E_g^N = {-3\over {5 - n}}
{G M^2 \over R} \e where $\gamma_p \equiv 1 + 1/n$ and $G$ is the
gravitational constant. Note that, in general, $\gamma \neq
\gamma_p$ and only if the fluid is considered to undergo adiabatic
change, one would have $\gamma = \gamma_p$. Further, as we would
see, all contraction processes are expected to be accompanied by
emission of radiation, and they must be non-adiabatic in a strict
sense. It may be also mentioned that, in the Newtonian case, the
fluid density appearing in Eq.(10)
 essentially means rest mass  density: $\rho = \rho_0$.

If one differentiates Eq.(8), for slow contraction, one will have \b
{dU \over dt} ={-1 \over3(\gamma -1)} {d E_g^N \over dt} \e Also,
from Eq.(11), we see that \b {d E_g^N \over dt} = {3\over 5 - n} {G
M^2 \over R^2} ~{\dot R} \e Since $\dot {R} < 0$ for contraction,
while the value of $E_g^N$ decreases during contraction its absolute
value $|E_g^N|$ increases. From Eq.(12) , we find that, as $|E_g^N|$
increases during such contraction, so does $U$. However, the amount
of total gravitational energy released by the contraction, $|d
E_g^N|$, is not fully accounted for by the gain in the value of $U$:
\b dU ={1 \over3(\gamma -1)} |d E_g^N| < |d E_g^N| \e For overall
energy conservation, it is therefore necessary that the rest of the
energy gain \b \l[1 -{1 \over3(\gamma -1)}\r] ~|d E_g^N| =  {3
\gamma -1 \over3(\gamma -1)} |~d E_g^N| =d Q \e is radiated away by
the system. This could have been found directly by differentiating
Eq.(9): \b {d E_N\over dt}  = {3 \gamma -4 \over3(\gamma -1)} ~{d
E_g^N \over dt} \e Using Eq.(13) into above Eq., we find that \b {d
E_N\over dt}  = {3 \gamma -4 \over3(\gamma -1)} ~{3\over 5-n}~{G
M^2\over R^2}~ {\dot R} <0 \e

In case of a gas confined in a laboratory by physical inclosure,
$p_b >0$. One can also imagine the physical inclosure to be a
perfect insulator and one may conceive of a radiationless adiabatic
contraction for arbitrary $\gamma \ge 4/3$. But in an astrophysical
context, there is neither any physical inclosure nor  any perfect
insulating surrounding. Hence, it appears from Eq.(15) that a
strictly adiabatic contraction ($d Q=0$) would be possible only for
the idealized case of $\gamma =4/3$. And Eq.(9) would show that, in
such a case, one would already have $E_N =0$, i.e., they system
would be unbound.  In reality, one has $\gamma =4/3$ only for pure
radiation or when the energy of the particles per unit rest mass
$E^* = \infty$, which is possible only for a singular situation in
case the ``gas'' is not already a pure radiation\cite{6}.

Thus, Eq.(17) shows that the total (Newtonian) energy of the system
decreases for contraction and it could be so only if the system
radiates appropriate amount of energy. Since $U$ increases, the
fluid become hotter while it radiates ($d Q <0$). Therefore a
self-gravitating fluid has a {\em negative specific heat} and this
fact is well known. Note that this result follows from the Newtonian
HK process and does not depend on the details of either the physical
properties of the fluid or the collapse process.

Since the above result is of generic nature, it is expected to be
qualitatively valid even in case of strong gravity. In fact, even
after the introduction of General Relativity (GR) into astrophysics,
the idea that gravitational contraction must result into radiation
output is naturally and justifiably used\cite{1}.  While considering
the configuration of static fluid spheres in GR, Buchdahl\cite{7}
posed the question whether the amount total radiation emitted by
gravitational contraction can {\em even exceed} the initial value of
$E= Mc^2$ itself. The answer was in the negative. Yet, for Einstein
gravity {\em no exact counterpart of Eqs.(12-17)  exists}. Thus, we
cannot assert, as a principle, that self-gravitating matter has
``negative specific heat'' in Einstein gravity too. And we want to
address  this precise aspect in the present paper.

\section{Static Fluids in General Relativity}
Let us consider a static self-gravitating fluid sphere
described by the metric
\begin{equation}
ds^2 = A^2(r) dt^2 - B^2(r) dr^2 - r^2 (d\theta^2 + \sin^2 \theta d\phi^2)
\end{equation}
Here we have taken $G=c=1$. Recall that one often uses the symbols
$A^2 = e^\nu = g_{00}$ and $B^2 = e^\lambda = -g_{rr}$. The radial
coordinate $r$ is the luminosity distance and as before, the
coordinate volume element is $dV = 4 \pi r^2 dr$. The proper volume
element however is \b d {\cal V} = B(r) ~dV = 4 \pi r^2~B(r)~ dr \e
Note that since $B(r) >1$, $d {\cal V} > dV$, and this might be seen
as the effect of ``stretching'' of space by gravity. The energy
momentum tensor of the body in mixed tensor form is
 \begin{equation}
T^i_k = (p+\rho) u^i u_k +p g^i_k
\end{equation}
where $u^i$ is the fluid 4-velocity,
$\rho$ is the total mass energy density (excluding contribution due to self-gravitation),
and $p$ is the isotropic pressure.

The total mass energy density of a cold fluid (excluding  negative
self-gravitational energy) is
\begin{equation}
\rho = \rho_0 + e
\end{equation}
where $\rho_0 = m_N n$ is the proper rest mass energy density, $m_N$
is nucleon rest mass, $n$ is nucleon proper number density (not to
be confused with polytropic index), $e$ is the internal energy
density, and $g^i_k$ is the metric tensor. The explicit form of
$B(r)$ is\cite{8}
\begin{equation}
B(r) = [1- 2M(r)/r]^{-1/2} = (1- 2m/r)^{-1/2}
\end{equation}
where
\b
 M(r) = m = \int_0^r 4\pi \rho ~r^2 dr
 \e
The total energy of the system as perceived by a distant inertial observer, $S_\infty$, i.e.,
the gravitational mass of the fluid is
\begin{equation}
M = \int_0^R ~\rho ~dV =\int_0^R (\rho/B)~ d{\cal V}
\end{equation}
This total energy is also known as gravitational mass or {\sl
Schwarzschild mass} of the fluid although actually this should have
been named by the name of Hilbert. Note that
\begin{equation}
M =\int_0^R (\rho/B)~ d{\cal V}\neq \int_0^R \rho ~d{\cal V}
\end{equation}
as one might have expected. The reason for this is that total mass
energy includes not only local contributions from $\rho$ but also
the {\em negative} global contribution of self-gravitational energy.
It is this negative latter contribution which reduces the effective
net proper energy density from $\rho$ to $\rho/B(r)$.

On the other hand, the total {\em proper} energy content of
 the sphere, by excluding any negative self energy contribution , i.e., the energy obtained by merely adding
 individual local energy packets,
   is
\begin{equation}
M_{proper} = \int_0^R \rho ~d{\cal V}
\end{equation}

The {\em conventional} definition GR self- gravitational  energy of
the body is\cite{7,8}
\begin{equation}
E_G =  M - M_{proper} = \int_0^R (1- \sqrt{-g_{rr}}) ~\rho~ dV
\end{equation}
Since $B(r) >1$ in the presence of mass energy, $E_G$ is a -ve
quantity, as is expected.  Note however that the $E_G$ defined by
Eq.(27) is the sum of appropriate local (proper) quantities somewhat
like the definition of $M_{proper}$ in Eq.(26); and is not defined
with respect to the inertial observer $S_\infty$. Note also that, in
contrast, gravitational mass $M$ is indeed the mass-energy measured
by $S_\infty$.

The {\em proper} rest mass energy of the fluid is \b M_0 = \int
\rho_0 ~ d{\cal V} \e  If there are no initial anti-baryons or
anti-leptons, $M_0 = m_N ~ N$, where $N$ is the total number of
baryons, and is therefore a conserved quantity.

The {\em proper} internal energy of the fluid is
\b
 U = \int e~
d{\cal V} \e As before, note that, neither $M_0$ nor $U$ is defined
with respect to $S_\infty$. By using Eqs.(20-28), it can be verified
that \b
 E_G + U = M - M_0
 \e
 From the above equation, it may {\em appear} that the GR equivalent of $E_N$, the binding energy, is
\b E_{GR} = M - M_0 = E_G + U \e Suppose the gas was initially
infinitely  dispersed to infinity. Further, let the gas molecules be
rest with respect to $S_\infty$. Under such a case, the initial mass
energy of the cloud is just the rest mass energy: \b M (t=0) = M_0
\e And if the contraction of the cloud into a finite size would
indeed release energy, one must have $ M - M_0 <0$. But unlike the
Newtonian case, as noted by Tooper\cite{9}, it is not apparent from
the definitions of $E_G$ and $U$ that $E_{GR} = E_G + U <0$! This is
so because  while in the Newtonian case, $E_g^N$ and $U$ are related
through Eq.(7), {\em there is no known relationship} between $E_G$
and $U$ in the GR case. Further there might be examples when the
occurrence of a negative $E_{GR}$ ``{\sl is not a sufficient
condition for instability of the system against an expansion to
infinity}''\cite{9}. This means that $E_{GR}$ may not be the correct
GR equivalent of a ``binding energy''. Had $E_{GR}$ been the true GR
equivalent of $E_N$, probably, one would have had
 equations similar to (8-9) involving $E_G$, $U$ and $E_{GR}$. But no such equations exist.
Thus, although, intuitively, one expects GR contraction to release radiation energy,
one really cannot show it unlike the Newtonian case developed in previous section.

In Eq.(30), we may note that, while, $E_G$, $U$ and $M_0$ {\em are
not} defined with respect to (w.r.t.) $S_\infty$, $M$, on the other
hand {\em is defined} w.r.t. $S_\infty$. And this may be the
fundamental reason that $E_{GR}$ defined in Eq.(31) may not be the
true ``binding energy'' of the fluid.

\section{ Another Definition of Fluid Mass}
For any stationary gravitational field, total four momentum of
matter plus gravitational field is conserved and independent of the
coordinate system used\cite{10,11}:
\begin{equation}
P^i = \int (T^{i0} + t^{i0}) ~dV
\end{equation}
where $t^{ik}$ is the energy momentum pseudo-tensor associated with
the gravitational field. Further,  the inertial mass (same as
gravitational mass),
 i.e, the time component
of the
  4-momentum of any given body in GR can be expressed
  as\cite{10,11,12}
 \begin{equation}
M = \int_0^\infty (T^0_0 -T^1_1 -T^2_2 -T^3_3) ~\sqrt {-g}~ d^3 x
 \end{equation}
 where
  \b
  g = - r^4 A(r) B(r) \sin^2\theta
  \e
   is
the determinant of the metric tensor $g_{ik}$ and $ d^3x = dr~d\theta~d\phi$. Since
\b
T^1_1 = T^2_2 =T^3_3 = -p; \qquad T^0_0 = \rho
\e
  it follows that\cite{11,12}
\begin{equation}
M = \int_0^\infty (\rho + 3p) A(r) B(r) dV = \int_0^\infty (\rho + 3p) A(r) d{\cal V}
\end{equation}
When the body is {\em bound} and $p_b =\rho_b =0$ for $r \ge R$,
then, the foregoing integrals shrink to\cite{11,12}
\begin{equation}
M = \int_0^R (\rho + 3p) A(r) B(r) dV = \int_0^R (\rho + 3p) A(r) d{\cal V}
\end{equation}

Although our result would not depend on splitting of the foregoing
equation, (since we would simply equate the ``total'' expression of
``inertial'' and ``gravitational'' mass), we might nevertheless do
so
\begin{equation}
M =  \int_0^R \rho ~ A(r)~ d{\cal V} + \int_0^R 3~ p ~ A(r)~ d{\cal V}
\end{equation}
for the sake of obtaining  physical insight.
\subsection{Global Definitions}
The Newtonian virial theorem (4) is essentially a statement of
energy conservation involving negative self-gravitational  energy
and positive thermodynamic energy of the fluid. In the Newtonian
case, there exists  global inertial frames and a statement of energy
conservation can be made in a trivial way. But in GR, even for this
simplest case of a static fluid sphere, there is no global inertial
frame. Thus the exercise of having global definitions of related
energies and to enact their conservation is a highly non-trivial
task.
 As is well known, for stationary systems, however global energy can be defined
in a meaningful way for asymptotically flat spacetimes. Further,
when the energy is defined with reference to an observer at a
spatial infinity ($S_\infty$),  we obtain the so-called ADM
Mass\cite{13}. Also global energy conservation can be meaningfully
defined only with reference to $S_\infty$.

Note that $E_G$ and $U$ are essentially summation over {\em local}
appropriate values and {\em not} over the corresponding quantities
measured by $S_\infty$. It may be mentioned that, if {\em any}
locally measured energy is $  \epsilon$, then the energy measured by
the far away inertial observer is the redshifted quantity

\b \epsilon_\infty = \sqrt{g_{00}}~\epsilon = A(r) ~\epsilon
 \e
Accordingly,  the total mass energy content of the fluid, excluding
any self-energy,
 as measured by an inertial frame such as the distant observer, $S_\infty$ is
{\em different} from $M_{proper}$ and is {\em given by}\cite{8} (Eq.
11.1.19):

\begin{equation}
M_{matter} = \int_0^R A(r)~ \rho ~d{\cal V} =\int A(r)~ B(r) ~\rho ~dV
\end{equation}
Physically this means that local energy content in a given cell $\rho d{\cal V}$
is measured ({\em redshifted}) as $A(r) \rho d{\cal V}$ by the inertial observer $S_\infty$.

In fact, the {\em inertial} observer $S_\infty$ would see the rest
mass energy, i.e., the proper energy, of a nucleon too to be reduced
by the same factor  $ \sqrt{g_{00}}$. Hence, when the body is finite
and not dispersed to infinity, the total rest mass energy of the
body, as reckoned by the inertial observer $S_\infty$, is \b {\tilde
M_0} = \int_0^R \rho_0~A(r)~ d{\cal V} \e Similarly, the {\em
redshifted} global internal energy of the fluid as measured by
$S_\infty$ is
\begin{equation}
{\tilde U} = \int_0^R  e(r) A(r) ~d{\cal V}
\end{equation}

 We can, now, quickly identify the 1st term on the RHS of Eq.(39) as $M_{matter}$.
 Again bear in mind the fact that our eventual result would not
 depend on such identification or splitting because it would be
 obtained by equating the ``total'' expressions for inertial and
 gravitational masses.

Further, we see that, the 2nd term on the RHS of Eq.(39) is the
global energy associated with pressure as perceived by $S_\infty$. Accordingly, we rewrite Eq.(39)
as:
\begin{equation}
M =  M_{matter} + M_{pressure}
\end{equation}
Again recall that $M$ too is defined only with respect to
$S_\infty$. Having done this splitting, we are in a position to
obtain the GR  Virial Theorem, which is essentially an accounting of
global energies involved in the problem. And since global energies,
in GR, can be defined only w.r.t. the inertial observer $S_\infty$,
all relevant integrals must be defined w.r.t. the same observer. And
this is what we have just done.
\section{ GR Helmholtz Kelvin Process}
Let us simply transpose Eq.(39) (irrespective of its splitting) as

\begin{equation}
 \int_0^R \rho ~ A(r)~ d{\cal V} -M + \int_0^R  3p~ A(r) d{\cal V} =0
\end{equation}
to reexpress  as
\begin{equation}
{\tilde E_g} + \int_0^R  3p~ A(r)~ d{\cal V} = 0
\end{equation}
Or,
\begin{equation}
{\tilde E_g} +
 \int 3p ~\sqrt{-g_{00}~ g_{rr}} ~dV = 0
\end{equation}
where
\begin{equation}
{\tilde E_{g}} = M- M_{matter}  =  \int (AB-1)~ \rho ~dV
\end{equation}
Or else,
\begin{equation}
{\tilde E_{g}}  =\int (\sqrt{-g_{00} g_{rr}} -1) ~\rho ~dV
\end{equation}
Since $AB <1$ in the  presence of mass-energy, we have ${\tilde E}_g
<0$ as is expected. Clearly, we have, obtained, now an equation
similar to (4). It appears then that the above defined ${\tilde
E_g}$ rather than the previously defined $E_G$ is the true measure
of self-gravitational energy as perceived by an inertial observer
$S_\infty$. This is so because $E_G$ is not defined with respect to
$S_\infty$, the accountant for global energy. In contrast, both the
components of ${\tilde E}_g$, namely, $M$ and $M_{matter}$ are
defined w.r.t. $S_\infty$.

 Further, recall that, in GR, the effect of ``gravitational potential'' is
conveyed by $g_{00}$. But $E_G$ (Eq.[27]) {\em does not contain} $g_{00}$ at all.
  In contrast, ${\tilde E_g}$ indeed involves gravitational potential term $g_{00} =A^2$ (Eq.[49]).

What would be the value of true global GR self-gravitational energy ${\tilde E_{g}}$ in the Newtonian limit?

To see this we consider a sphere with $\rho=$ constant for
which\cite{8}
\begin{equation}
A(r) = {1\over 2}[ 3(1- 2M/R)^{1/2} - B(r)^{-1}]
\end{equation}
Using Eq.(22) in (50), we further see that
\begin{equation}
A(r) B(r) = {1\over 2}[ 3(1- 2M/R)^{1/2} B(r)  - 1]
\end{equation}
Again using Eq.(22) in (51), we obtain
\begin{equation}
A(r) B(r)
 ={1\over 2}[ 3(1- 2M/R)^{1/2} (1-2m/r)^{-1/2}   -1]
\end{equation}
 Now if we proceed to linearized gravity limit with $M/R \ll 1$ and $m/r\ll 1$, we will have
\begin{equation}
A(r) B(r) = 1-  {3\over 2}(m/r- M/R)]
\end{equation}
Using Eq.(52) in (48),  we see that
\begin{equation}
{\tilde E_g} =  -{3\over 2}\int (M/R - m/r) \rho ~dV
\end{equation}
When we carry out this above integration with $\rho =constant$, we obtain
\begin{equation}
{\tilde E_g} =  {-3\over 5} {M^2\over R}
\end{equation}
Therefore, in the Newtonian limit, ${\tilde E_g} = E_G = E_g^N$, though in general
${\tilde E_g}$ and $E_g$ are different.

Now, using the thermodynamical relation (5) and Eq.(43) in Eq.(46), as before,  we will have
\begin{equation}
{\tilde E_g} + 3 (\gamma -1){\tilde U} =0
\end{equation}
By direct comparison with Eqs.(4) and (7), we can easily identify
Eqs.(48) and (56) as the appropriate GR version of static scalar
virial theorem. Note that Eqs.(46) and (56) naturally reduce to
their Newtonian forms, Eqs.(1) and (7) for sufficiently weak gravity
with $g_{00} \approx -g_{rr} \approx 1$. Thus we may interpret the
existence of the Newtonian virial theorem too as due to equivalence
of ``gravitational mass'' and ``inertial mass''.

From Eq.(56), we obtain
\begin{equation}
{\tilde U} = {-1\over 3(\gamma -1)} {\tilde E_g}
\end{equation}
If the fluid undergoes quasistatic contraction and $\Delta$ denotes the
associated changes in relevant quantities, then we will have

\begin{equation}
\Delta {\tilde U} = {-1\over 3(\gamma -1)} \Delta {\tilde E_g}= {+1\over 3(\gamma -1)} |\Delta {\tilde E_g}|
\end{equation}
Here we have used the fact that since ${\tilde E}_g <0$ it  must decrease
 for  contraction.
 Thus as is expected,
the internal energy of the fluid must increase for gravitational contraction.
If the appropriately averaged value of $A(r) = \bar {g_{00}}$ during this contraction,
the increase in proper internal energy would be
\begin{equation}
\Delta  U = {\Delta {\tilde U}\over {\bar g_{00}}}  = {1\over 3(\gamma -1)}
 {|\Delta E_g | \over {\bar g_{00}}}  >0
\end{equation}
Since $g_{00} <1$ in the presence of gravity, this means that, the {\em rate of increase
in proper internal energy would be  higher than in the corresponding Newtonian case}.

As we look back at  Eq.(58),  the increase in the value of $|{\tilde
E}_g|$, namely $|\Delta {\tilde E}_g|$
 is not fully accounted for by the increase in the value of ${\tilde U}$. Note that
 in the absence
  of initial antiparticles,
  the  contribution of rest mass-energy is unaffected during the process. Therefore, for the sake of  global energy conservation,
as reckoned by the inertial observer $S_\infty$, the fluid {\em must} radiate out
an amount of energy $+\Delta Q$ given by
\begin{equation}
\Delta Q = \l[1 -{1\over 3(\gamma -1)}\r]~ |\Delta {\tilde E_g}|
 = {3 \gamma -4 \over 3(\gamma -1)} |\Delta {\tilde E_g}|
\end{equation}
in order to be able to contract.
 Note, as before, that $\gamma > 4/3$ in a strict sense, as long as particle momenta are finite. To see that  in the GR context too, that $\gamma = 4/3$ implies singular situation,
 look at the {\em Ist row of  Table I} of \cite{14}
 which shows that in this case again $R_{max}=0$, where $R_{max}$ is the maximum possible
 radius of the configuration. Further, for $\gamma =4/3$, the next
 entry in the same row shows that
  $R_0 =0$  where
  $R_0= 2GM/c^2$ is the Schwarzschild radius. This implies that for
  $\gamma = 4/3$, one has $R_0 =0$.
  This latter fact implies
 that even in GR, {\em the total mass energy} $Mc^2 =0$
  just as $E_N =0$ for $\gamma = 4/3$ (Eq.[9]). Occurrence of
  $R_{max} =0$ implies a  fluid sphere that has {\em collapsed to a singular
  point}. And as per Ref.(14), {\em the singular configuration then would have zero mass
  energy}.  Incidentally, in the classic paper\cite{13}, Arnowitt, Deser and Misner too
  found that a neutral ``point particle'' has zero ``clothed mass''.
  Then Chandrasekhar's exercise\cite{14}, in addition, suggests that {\em if the fluid
  would attain such a singular state, the value of} $\gamma \to 4/3$.
  This is also in perfect agreement with the notion that a singular
  state should be infinitely hot with complete domination of
  radiation energy over rest mass energy\cite{6}.

 In fact, in one would misconceive of a situation with $\gamma <
 4/3$, irrespective of
  whether it is a Newtonian  or a GR case, one would have to ensure
  injection of energy into the system to let it  collapse. This would mean that in the absence of external injection of energy,
 the system would not contract/evolve at all in defiance of basic tenet of global gravitation.

Thus, one must indeed have $\gamma >4/3$ and,  in a strict sense,
{\em there cannot be any adiabatic gravitational contraction}.
Consequently, all strictly adiabatic gravitational collapse studies
are of only academic interest. Thermodynamically, the global
specific heat of the contracting fluid is negative because $dQ <0$
while the temperature and internal energy of the fluid increase.
\section{ General Relativistic Binding Energy}
Eq.(60) suggests that, we might isolate a quantity
\begin{equation}
{\tilde E}  = {3 \gamma -4 \over 3(\gamma -1)} ~{\tilde E_g}
\end{equation}
as  the total energy of the system excluding any rest mass
contribution. This is actually the ``Binding Energy'' of the
gravitating system, defined by a given distribution of mass-energy.

 By using Eq.(57), it is seen that
\begin{equation}
{\tilde E}  =  {\tilde E_g} + {\tilde U}
\end{equation}

Further using Eqs.(39), (42), (43) and (48), it also transpires that
\begin{equation}
{\tilde E}  =  {\tilde E_g} + {\tilde U} = M - {\tilde M_0}
\end{equation}
We may see that, {\em unlike in Eq.(30)}, all the quantities
involved in Eq.(63) are defined w.r.t. $S_\infty$ and which suggests
that $\tilde{E}$ is indeed the true binding energy of the fluid. The
reader is again requested here to appreciate the subtle point why
the true binding energy of the fluid is given by Eq.(63) rather than
by Eq.(31).:

Though $M_0$ is a conserved quantity, once the fluid is contracted
into a finite size, it gets dissociated from the inertial frame
$S_\infty$. And $S_\infty$, who is doing the global energy
accounting, sees the locally defined rest mass energy to be
redshifted to ${\tilde M_0}$ rather than as $M_0$. On the other
hand, the total mass-energy of the fluid, again defined w.r.t.
$S_\infty$ is $M$. Therefore, the global binding energy of the
fluid, i.e., the difference between the total mass energy and rest
mass energy as seen by the same inertial observer $S_\infty$ is $M -
{\tilde M_0}$. Note that the existence of equations (62-63) does not
depend on such interpretations because they, in any case, crept up
spontaneously.

Although, in a Newtonian case, the notion of a binding energy always
existed, to the best of our knowledge, such a notion was never
before properly derived in the GR context. A relativistic ``bound
system'' may thus be defined as one having ${\tilde E} <0$, and an
``unbound system'' will have ${\tilde E} >0$.

 While, in the Newtonian case, ``Total Energy'' is the binding energy $E_N$,
 in GR, total global energy, as measured by $S_\infty$, always is $E= Mc^2$.

\section{Discussions}
The important idea of Helmholtz and Kelvin, developed in the 19th
century, that gravitational contraction should both raise the
internal energy and cause the fluid to radiate was always expected
to be valid irrespective of the strength of the gravity. However,
the original derivation to this effect was made in the framework of
extremely weak, i.e., Newtonian gravity.  We showed here that even
for arbitrary, strong gravity, this process indeed remains valid. In
fact, as shown by Eq.(59), the process becomes even {\em more}
effective compared to the Newtonian case as gravity increases.
Pictorially, we may think that stronger gravity churns out more
radiation from matter even in the absence of chemical or
thermonuclear  energy sources. A similar conclusion is supported by
a recent work which shows that the ratio of radiation energy density
to rest mass energy density of a self-luminous contracting object is
proportional to its surface redshift $z$ \cite{6}. When $z \ll 1$,
the self-gravitating contracting object is ``matter dominated'',
i.e., $ \rho_0 \gg \rho_r$, but when, $z \gg 1$, the object becomes
radiation dominated : $\rho_r \gg \rho_0$ like the very early
Universe\cite{6}. The present study provided an additional
explanation for this result. These studies show that the actual fate
of radiative physical gravitational collapse could be radically
different from traditional pictures of continued gravitational
collapse inspired by the pressureless dust collapse  where a Black
Hole (BH) or a Naked Singularity is catastrophically formed in a
finite comoving proper time. Traditional GR collapse studies are
usually done  by (i) assuming dust models with $p\equiv 0$ even when
the fluid is supposed to have collapsed to singularity, or (ii)
considering pressure but neglecting all heat transport, $dQ \equiv
0$. But as shown by Eqs. (58) and (60), $dQ=0$ implies  (a) $d
{\tilde U} =0$ and (b) $d{\tilde E_g} =0$. The condition (a) is
satisfied only for dust and thus despite a formal consideration of
existence of pressure, in the context of collapse,  a fluid
satisfying condition (a) becomes similar to  a pressureless,
internal energyless dust. The condition (b) is not satisfied even
for a dust unless it has $M=U=E_g=fixed=0$ at the beginning of the
collapse. But no isolated fluid with finite size can have $M=0$ (the
Universe may, however, have $M=0$ even being of finite or infinite
extent). In several numerical studies of supposed radiative
collapse, one implicitly or explicitly assumes $ Q \ll M_0 c^2$. At
the advanced stage of collapse, this assumption fails\cite{6} and
such cases effectively become similar to case (ii) of adiabatic
continued collapse valid for $M=0$.

In contrast, in a breakthrough research on physical gravitational
collapse, Herrera and Santos\cite{1} have shown that the force
exerted on the collapsing fluid by the outward propagating
heat/radiation may stall the continued collapse process and
formation of either a finite mass BH or Naked Singularity may be
averted. Herrera, Di Prisco and Barreto\cite{1} have successfully
made a numerical model of continued collapse to substantiate this
pathbreaking idea. Such ideas are consistent with the model
independent generic studies\cite{6} which show that continued
catastrophic collapse indeed degenerates into a radiation pressure
supported hot quasistatic state called eternally collapsing objects
because of outward force due to collapse generated radiation at
extremely deep gravitational potential wells, $z \gg 1$ where $z$ is
the surface gravitational redshift of the collapsing object. It is
because of the resultant reduction in the value of $M$ due to
continuous radiation outpour that no  apparent horizon or event
horizon  is formed until $M=R =0$\cite{6}.

Finally, the GR definition of ``binding energy'' of a static fluid is
$M - {\tilde M_0}$
rather than $M- M_0$.

Newtonian HK process is a direct sequel of static Newtonian virial
theorem. Similarly, we needed to derive the exact GR version of the
static virial theorem.  This GR virial theorem  involved globally
defined quantities measured w.r.t. the {\em same} inertial observer
$S_\infty$. In general, the notion of global energy is far from
transparent and unique in GR. For example, for non-static and
non-spherically symmetric systems or charged systems, there could be
various notions of ``energy'' and ``mass''; to appreciate this one
may have a look at a recent long review paper \cite{15}. However the
present paper must not be confused with such studies. The aim of
this paper was not at all to define any new definition of either
``mass'' or ``global energy'' from any preferred theoretical
perspective, correct or incorrect. This is so because, unlike a
generic case, the definition of ``global mass energy'' of a
chargeless  static spherically symmetric fluid (measured by
$S_\infty$) is very well known since long\cite{10,12}. And we just
appealed to the Principle of Equivalence to equate the {\em already
well known} expressions for ``gravitational mass'' (Schwarzshild
mass) and ``inertial mass'' - the time component of linear
4-momentum. It is this simple operation which yielded the GR virial
theorem and GR HK mechanism (Eqns. [12-17]). It is the same
principle of equivalence which demanded that, from energy
conservation considerations, all the relevant globally defined
energies are defined w.r.t. the {\em unique inertial} frame
$S_\infty$ rather than w.r.t. a series of (infinite) proper frames.
To the best of our knowledge, this is the maiden derivation of HK
process using GR. And this is also the maiden proper GR explanation
for the intuitive notion that a self-gravitating object has
effective global ``negative specific heat'' in Einstein gravity too.

 For further appreciation of the ``global quantities'' involving
 $\sqrt{g_{00}}$ in our study which arose spontaneously and not imposed
 from new theoretical perspective, we recall that the ``Poisson's
 Equation'' in GR has the form\cite{16}
 \b
 \nabla^2 \sqrt{g_{00}} = 4 \pi G \sqrt{g_{00}} ~ (\rho + 3p)
 \e

 And only when one moves to weak gravity with $GM/r \ll 1$ and $\sqrt{g_{00}}
 \approx
 1 - GM/r \approx 1$, one
 obtains the more familiar form

 \b
 \nabla^2 \phi  = 4 \pi G  ~ (\rho + 3p)
 \e
 where the weak ``gravitational potential $\phi \sim  G M/r$. Of
 course, Eq.(64) would also degenerate into Eq.(65) in a {\em local}
 free falling frame where $g_{00} =1$. But for fluids having finite
 pressure there cannot be any such {\em global} free falling frame
 and therefore Eq.(65), in the present context, can be recovered
 only for weak gravity (in any case we are dealing with a {\sl
 static} fluid. Formally a strict adiabatic collapse is possible only for a pressureless ``dust''
 with $p =U \equiv 0$, though, physically,  $M=0$ in such a case.

Since virial theorem  is important for the study of compact objects
and gravitational contraction, the {\em exact} relativistic virial
theorem obtained here  could be useful for relativistic
astrophysics, either now or in future.

\section{Acknowledgements} The author thanks Felix Aharonian
for encouragement. MPI fur Kernphysik, Heidelberg is thanked for the
kind   invitation and hospitality. Both the anonymous referees are
also thanked for unprejudiced fair assessment of the manuscript and
some minor suggestions. Finally, the Editorial Staff members of
Phys. Rev. D. are thanked for a speedy processing of this
manuscript.

\end{document}